\documentclass[preprint,showpacs,preprintnumbers,amsmath,amssymb]{revtex4}
\usepackage{graphicx}     
\usepackage{times} 
\usepackage{color}
\graphicspath{{.},{../Ps/},{$LOCALHOME/Obj/Sitdrop/Ps/}}%
\newcommand{\mylab}[1]{\label{#1}}
\begin{document}
\title{Liquid transport generated by a flashing field-induced wettability ratchet}
\author{Karin John}
\email{kjohn@spectro.ujf-grenoble.fr}
\affiliation{Laboratoire de Spectrom\'etrie Physique, Universit\'e Joseph
Fourier - Grenoble I, BP 87 - 38402 Saint-Martin-d'H\`eres, France}
\author{Uwe Thiele}
\email{thiele@mpipks-dresden.mpg.de}
\homepage{http://www.uwethiele.de}
\affiliation{Max-Planck-Institut
f\"ur Physik komplexer Systeme, N{\"o}thnitzer Str.\ 38, 
D-01187 Dresden, Germany}
\begin{abstract}
We develop and analyze a model for ratchet-driven macroscopic
transport of a continuous phase.  The transport relies on a
field-induced dewetting-spreading cycle of a liquid film with a
free surface based on a switchable, spatially
asymmetric, periodic interaction of the liquid-gas interface and
the substrate. The concept is exemplified
using an evolution equation for a dielectric liquid film 
under an inhomogeneous voltage. We analyse the influence of
the various phases of the ratchet cycle on the transport
properties. Conditions for maximal transport and the efficiency of
transport under load are discussed.
\end{abstract}
\maketitle
%
%
%

Brownian ratchets present a well established way to induce directed
motion of particles in spatially extended systems without global
gradients, i.e.\ without a globally broken spatial symmetry
\cite{Astu97,AsHa02,HMN05}.  Examples include colloidal particles
suspended in solution that move when exposed to a sawtooth dielectric
potential \cite{RSAP94} and the selective particle filter formed by a
micro-fabricated silicon membrane with asymmetrical bottleneck-like
pores under application of an oscillating pressure gradient
\cite{MaMu03}. Brownian ratchets are also thought to represent the
underlying mechanism of molecular motors that are responsible for the
active transport of molecules along filaments in biological cells
\cite{JAP97}.
Brownian transport is based on a principle first pointed out by Pierre
Curie \cite{Curi1894} stating that also a macroscopically symmetric
constellation may induce macroscopic transport if it exhibits local
asymmetries, e.g.\ a periodic asymmetric potential that varies on a
small length scale. However, the system has to be kept out of
equilibrium, for instance, by a chemical reaction \cite{JAP97}, an
oscillating pressure \cite{MaMu03} or electric potential
\cite{RSAP94}.

Ratchets are not only used to transport or filter discrete objects
like colloidal particles or macromolecules. A ratchet geometry may
also serve to induce a macroscopic transport of a continuous phase
using local gradients only. Experimentally investigated examples
include a secondary liquid flow triggered in {M}arangoni-{B\'e}nard
convection involving a solid substrate with asymmetric grooves
\cite{SISW03}.  On a similar substrate Leidenfrost drops perform a
directed motion \cite{QuAj06,Link06}. Micro-drops confined in
asymmetrically structured geometries move when vibrating the substrate
or applying an on/off electric field \cite{BTS02}.

\begin{figure}[tbh]
\includegraphics[width=0.9\hsize]{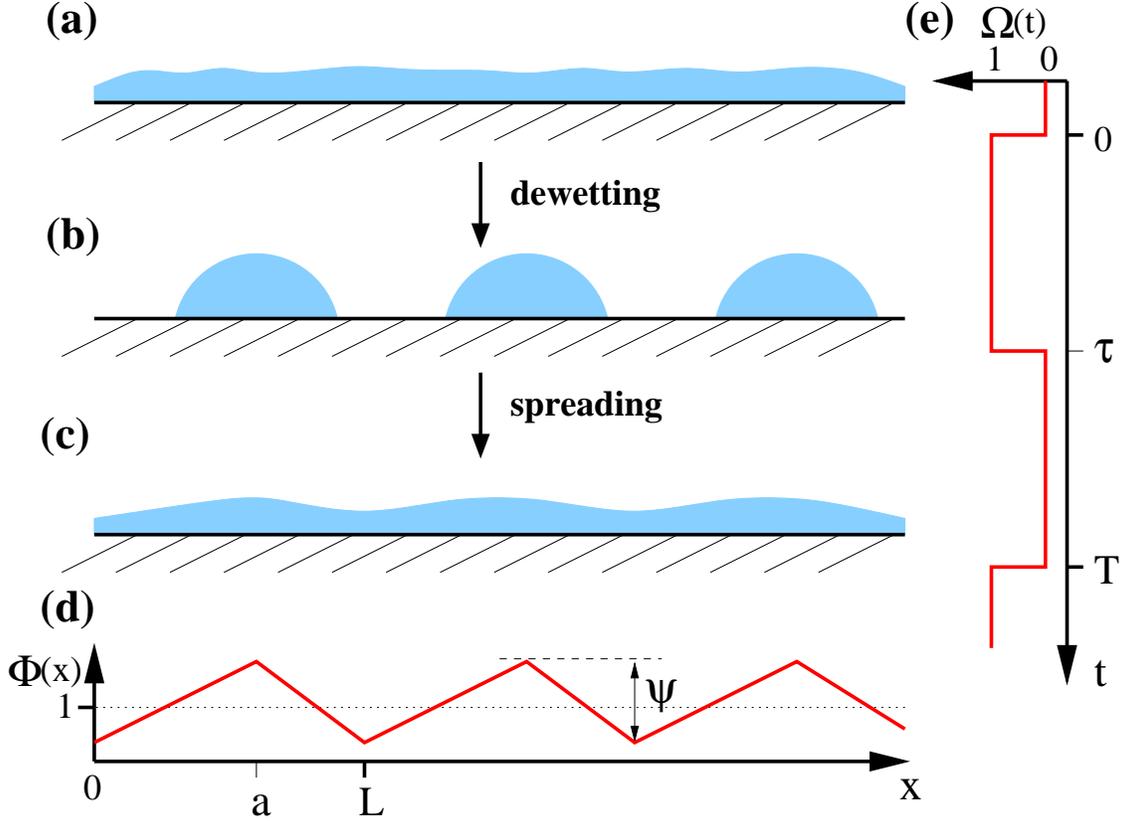}
\caption{
Panels (a) to (c) illustrate the working principle of a fluidic ratchet
based on a switchable wettability that causes dewetting-spreading cycles.
(d) illustrates the spatial asymmetric periodic voltage profile $\Phi(x)$ responsible
for the wettability pattern and (e) indicates the time dependence $\Omega(t)$ of 
the switching in relation to the dewetting and spreading phases in (a) to (c).
}
\mylab{fig1}
\end{figure}

In this Letter we propose and analyse a simple generic model for
ratchet driven free-surface flows resulting in the macroscopic
transport of a continuous phase. The driving flashing ratchet is based
on a switchable spatially periodic but asymmetric interaction of the
liquid-gas interface and the solid substrate like, for instance, a
field-induced wettability.  One can envision several ways to
experimentally realise the spatially inhomogeneous interaction and the
switching in time. A simple setup consists of a thin film of
dielectric liquid in a capacitor producing a sawtooth voltage profile
when switched on.  
The electro-dewetting provoked by the application of a homogeneous
electrical field is already used to structure thin films of liquid
polymers \cite{Lin01,STRS01,Mora03}.

An idealized electrical wettability ratchet 
(sketched in Fig.~\ref{fig1}) works as follows. We
consider a flat wetting film stable in the absence of an electric
field (Fig.~\ref{fig1}\,(a)).  Upon switching on an electric field at $t=0$
(Fig.~\ref{fig1}\,(e)) the film is destabilized by both, the overall
electric field and its local gradients (Fig.~\ref{fig1}\,(d)).  This
behavior is similar to dewetting on a heterogeneous substrate
\cite{KaSh01,TBBB03}. 
After a transient it results in the collection
of all the liquid in drops situated at the patches of maximal voltage 
(Fig.~\ref{fig1}\,(b)). After switching off the field
at $t=\tau$ (Fig.~\ref{fig1}\,(e)) the drops spread into a
homogeneous wetting layer (Fig.~\ref{fig1}\,(c)). 
Finally, at $t=T$ the field is switched on again and 
a new cycle starts. A typical realistic film dynamics during
one ratchet cycle is shown in Fig.~\ref{evol1}. 
For the geometry in Fig.~\ref{fig1}\,(d), with a voltage
profile shaped like a sawtooth skewed to the right, each cycle transports
liquid to the right.

The inhomogeneous electric field represents one out of several
possible choices for a switchable spatially asymmetric interaction of
liquid-gas interface and substrate. Others not pursued here include
switchable brushes \cite{Sido99}, heating \cite{ThKn04,BPT03} or the
switching of optical substrate properties, i.e.\ van der Waals
interactions. Note however, that although length and time scales and
the specific form of the resulting interaction potentials will differ,
in principle all these realizations can be mapped onto our model using
appropriate pressure terms corresponding to an effective
switchable 'wettability'.


In presenting the dynamical model we restrict our attention to a
two-dimensional system corresponding to a shallow open channel geometry 
and neglect the influence of the channel walls.
The interplay of field-driven
dewetting, liquid motion caused by local lateral gradients and spreading for
small scale systems, i.e.\ of wettability and capillarity dominated fluid flow,
is well described using an evolution equation for the film thickness profile
\cite{ODB97}. A derivation from the basic transport equations using
lubrication approximation gives
\begin{equation}
\partial_t\,h\,=\,\partial_x\,\left[\frac{h^3}{3\eta}\left(\partial_x\,p 
- f_{\mathrm{ext}}\right)\right]
\mylab{film}
\end{equation}
where $\gamma$ and $\eta$ are the surface tension and dynamic
viscosity of the liquid, respectively. $f_{\mathrm{ext}}$ denotes 
an external force in the direction of the x-axis. The change in time of the film
thickness profile is given by the divergence of the flow, expressed as
the product of a mobility and a pressure gradient and/or external force.
The velocity field within the film is fully determined by the film thickness
profile. The velocity parallel to the substrate is
$u(x,z)=\left(z^2/2-z h\right)\, \partial _x p$ and the vertical component
is obtained using continuity.
The pressure 
\begin{equation}
p\,=-\,\gamma\partial_{xx} h\,-\,\Pi(h,x,t)
\mylab{press}
\end{equation}
contains the curvature pressure $-\gamma\partial_{xx} h$ and the
disjoining pressure $\Pi(h,x,t)$.  The latter comprises the effective
interactions between the liquid-gas interface and the substrate, i.e.\
the wettability properties \cite{deGe85,Isra92,ODB97}.  Note, that the
lubrication approximation can be formally applied to systems involving
small surface slopes only.
However, even for
partially wetting systems with large contact angles, the lubrication
approximation predicts in most cases the correct qualitative behavior \cite{ODB97}. 

As model system we use a dielectric oil in a capacitor of gap width $d$. The
oil wets the lower plate and does not wet the upper plate corresponding to the
van der Waals disjoining pressure
\begin{equation}
\Pi_{vdW}=\frac{1}{6\pi}\left(\frac{A_l}{h^3}+\frac{A_u}{(d-h)^3}\right)
\end{equation} 
with the Hamaker constants $A_l>0$ and $A_u<0$.
A dielectric film in a capacitor with an applied voltage $U_0$ 
is subject to an electrical 'disjoining' pressure \cite{Lin01,MPBT05}
\begin{equation}
\Pi_{el} = \frac{1}{2}\varepsilon_0\varepsilon_1 U_0^2\,
\frac{(\varepsilon-1)}{[\varepsilon d + 
(1-\varepsilon)h]^2}.
\end{equation}
where $\varepsilon_0$ and $\varepsilon_1$ 
are the absolute and relative dielectric constant, respectively and
$\varepsilon=\varepsilon_1/\varepsilon_2$ denotes the ratio of the
relative dielectric constants of the liquid and the gas phase.  
The modulation of the electric field
is periodic in $x$, but breaks parity, i.e.\ there exists
no reflection symmetry in $x$.  This ratchet potential is periodically
switched on and off. In consequence $\Pi$ in (\ref{press}) takes the form
\begin{equation}
\Pi(h,x,t)=\Omega(t)\,\Phi(x)\,\Pi_{el}(h)+\Pi_{vdW}(h). \mylab{dp}
\end{equation}
where the temporal variation $\Omega(t)$ and spatial variation $\Phi(x)$ are
defined in Fig.~\ref{fig1}\,(a) and \ref{fig1}\,(b), respectively,
with $(1/L)\int_0^L\Phi(x)\,dx=1$.

\begin{figure}[tbh]
\includegraphics[width=0.8\hsize]{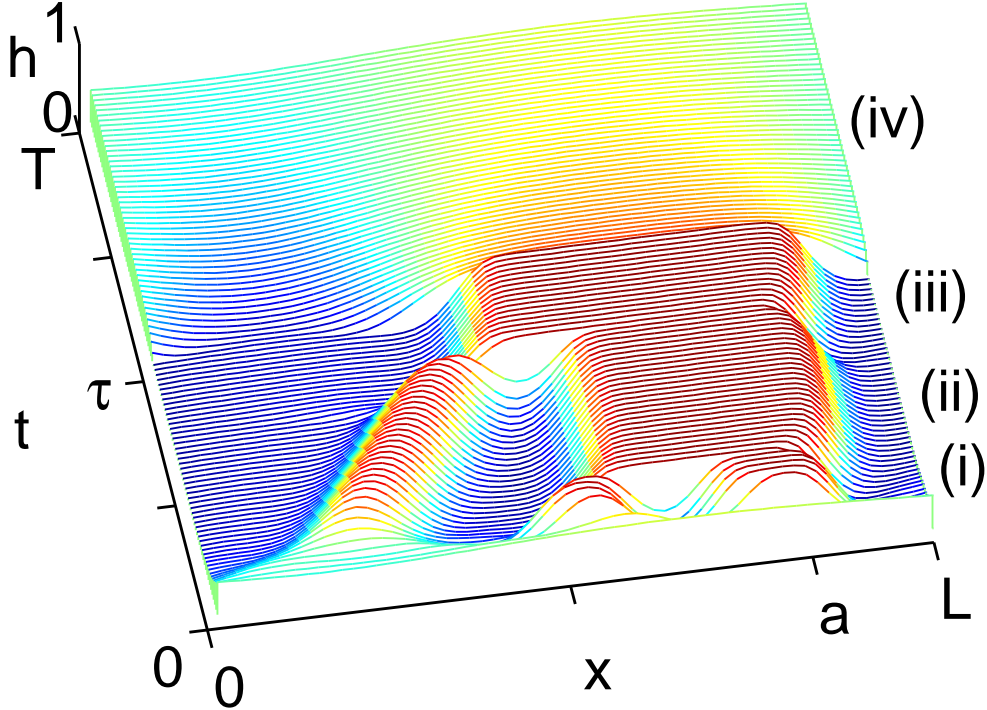}
\caption{Space-time plot of one temporal period 
of the evolution of the film thickness profile
for a film of dielectric liquid in a capacitor. Shown is
one spatial period. 
The different phases (i)-(iv) indicated to the right are explained in the main text. 
Parameters are $\bar{h}=0.5$, $\psi=0.5$, $L=32$, $\phi=5$, $T=5000$, 
$\omega=1$, $A=0.001$, $\varepsilon=2.5$ and $f_{\mathrm{ext}}=0$. The starting time is well after initial 
transients have decayed.
}
\mylab{evol1}
\end{figure}

To obtain a minimal set of parameters we introduce 
the scales $3\gamma\eta/d\kappa^2_{el}$, $\sqrt{\gamma d/\kappa_{el}}$, 
and $d$ for $t$, $x$,  and $h$, respectively.
The electrostatic 'spreading coefficient' is defined by
$\kappa_{el}=\varepsilon_0\varepsilon_1 U_0^2/2d^2$.  The resulting
dimensionless equations correspond to (\ref{film})-(\ref{dp}) with
$3\eta=\gamma=d=\varepsilon_0\varepsilon_1 U_0^2/2=1$. 
For simplicity we assume $A_u=-A_l$ and
obtain the dimensionless Hamaker constant $A=A_l/6\pi d^3\kappa_{el}$. All
results are given in terms of dimensionless quantities.

We characterize the ratchet using two measures, the asymmetry ratio
$\phi=(L-a)/a$ of the spatial variation $\Phi(x)$ and the flashing
ratio $\omega=\tau/(T-\tau)$ of the temporal switching
$\Omega(t)$. Zero net transport is expected for a symmetrical ratchet,
i.e.\ for $\phi=1$.
The resulting transport is quantified by the mean velocity $\bar{v}=\bar{j}/\bar{h}$ 
where $\bar{j}=(1/TL)\int_0^T dt\int_0^L dx\,j(x,t)$ is the mean flow along the
substrate and $j(x,t)=-h^3(\partial_x p - f_{\mathrm{ext}})$.

Figure~\ref{evol1} shows a typical example of the film evolution 
during one ratchet cycle allowing to distinguish four phases.
(i) When the ratchet is switched on the film is
nearly flat but rapidly evolves a surface instability with a wavelength given
approximately by the corresponding spinodal length \cite{ODB97}.
(ii) Next, the evolving profile coarsens accelerated by the gradients of the
ratchet potential. 
(iii) In an ideal situation only one drop remains, corresponding to
the equilibrium structure on the heterogeneous wettability pattern
produced by the ratchet potential \cite{TBBB03}.  
(iv) Finally, after switching off the ratchet the drop
starts to spread rapidly under the influence of van der Waals forces
until the next cycle starts.

\begin{figure}[tbh]
\includegraphics[width=0.8\hsize]{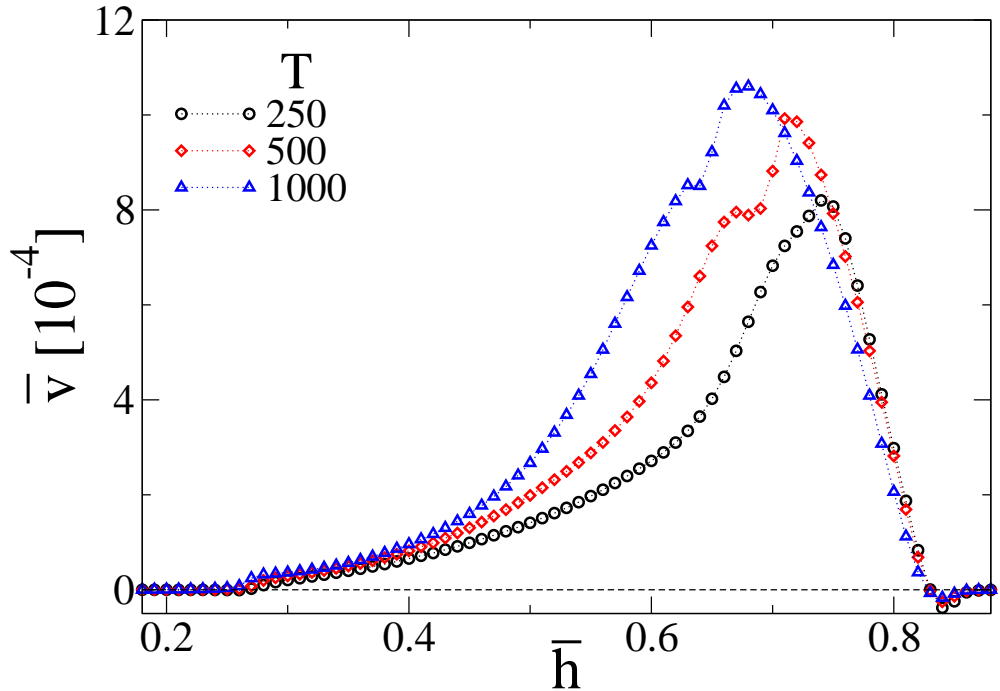}
\caption{Dependence of the mean velocity $\bar{v}$ on the liquid level in
  the capacitor, i.e.\ on the mean film thickness $\bar{h}$, for various
  flashing periods $T$ as indicated in the legend. The remaining
  parameters are as in Fig.~\ref{evol1}.  } \mylab{hdep}
\end{figure}

The competing influence of the various parameters allows to tune the
transport properties by adjusting the relative importance of the
various phases.  As shown in Fig.~\ref{hdep} the transport is
strongest for intermediate film thicknesses.  A film of intermediate
thickness in a homogeneous electric field ($\Phi(x)=\Omega(t)=1$)
dewets spinodally with a wavelength well below the used spatial period
of the ratchet. Therefore the ongoing coarsening interacts with the
flow induced by the spatial asymmetry and leads to a nontrivial
dependence of the mean velocity on the film thickness. For
very small or large film thicknesses the stabilizing van der Waals
terms are dominating and the mean velocity approaches rapidly zero. Note
also the flow reversal, albeit with a small mean velocity, for large
$\bar{h}\gtrsim0.85$.
Further calculations (not shown) demonstrate the monotonous increase
of the mean velocity with increasing asymmetry ratio $\phi$ or
amplitude $\psi$ of the ratchet potential for otherwise constant
parameters.

\begin{figure}[tbh]
\includegraphics[width=0.8\hsize]{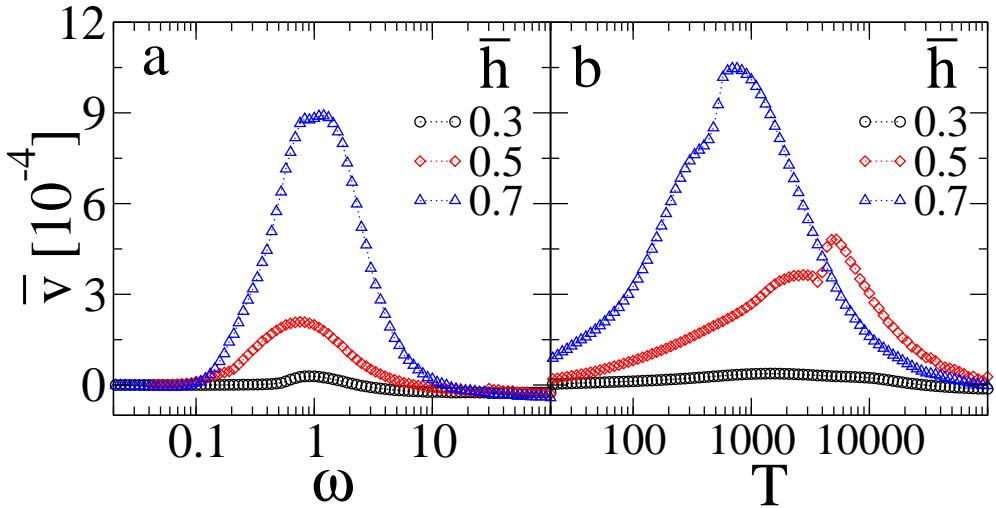}
\caption{Variation of the mean velocity $\bar{v}$ in dependence of (a) the
flashing ratio $\omega$ (for T=500) and (b) the flashing period $T$
(for $\omega=1$) for various film thicknesses as given in the
legends. The remaining parameters are as in Fig.~\ref{evol1}.  }
\mylab{gamandT}
\end{figure}

The flashing characteristics of the ratchet have a very pronounced
effect on the transport.  Figures~\ref{gamandT}\,(a) and (b) present
the non-monotonous dependencies of the flow on the flashing ratio
$\omega$ and period $T$, respectively.  For small $\omega$ the flow is
practically zero, since the time for dewetting is too short to trap a
considerable amount of liquid at the spots of high wettability.
Increasing $\omega$ increases the velocity until it reaches a maximum at
$\omega\approx{}1$. Beyond the maximum the flow decreases again
because less and less time remains for the spreading. For large
$\omega$ one observes a flow reversal.

The dependence of the velocity on the flashing period $T$ is similar but shows
around the flow maximum a particularly interesting non-monotonous behavior
that is due to coarsening.  For small
periods the fluid has neither enough time to dewet nor to spread and
the resulting mean transport is small. For large periods both
processes reach the respective equilibrium structure well before
the next switching, i.e.\ most time is spend waiting and
the mean velocity decreases approximately as $1/T$.

Assuming an ideal combination of ratchet properties and intrinsic
length and time scales, and approximating the intermediate drops at
the field maxima as point-like objects, all the liquid moves by
$a-L/2$ in one cycle of length $T$. For a film of mean
thickness $\bar{h}$ this yields a mean flow of
$\bar{j}_{ideal}=\bar{h}(a-L/2)/T$, i.e.\ the mean velocity is
$\bar{v}_{ideal}=(a-L/2)/T$. The maximal spatial asymmetry is
given by $a=L$, i.e.\ $\bar{j}_{max}=\bar{h}L/2T$ and
$\bar{v}_{max}=L/2T$.  In a real system, however, several effects keep
the mean velocity below the ideal value. For optimal parameters in
simulations we reach about 30\% of $\bar{v}_{ideal}$
(Fig.~\ref{gamandT}\,(b) for $\bar{h}=0.7$ and $T\approx10000$).

For practical applications transport against an external force is of crucial
importance.  In the reference system of a uniform substrate no work is
performed 
and all energy is lost via viscous dissipation. 
This is not the case for a thin film ratchet under load, i.e.\
$f_{\mathrm{ext}}\neq0$ in (\ref{film}). Switching off the ratchet,
$f_{\mathrm{ext}}<0$ induces a negative macroscopic flow.
Simple mechanical realizations of $f_{\mathrm{ext}}$ are an inclined substrate or
centrifugal forces yielding a constant $f_{\mathrm{ext}}$ independent of the film
thickness.  The flashing ratchet generates a positive flow, i.e.\ mechanical
work is performed against the external force.  The transport under load is
characterized by the (mechanical) energy transport efficiency
\begin{equation}
\nu_{\mathrm{eff}} = -\frac{\dot{W}}{\dot{Q}}
\end{equation} 
where 
\begin{equation}
\dot{W} =\frac{1}{TL}\int_0 ^L dx \int_0 ^T dt\, j \,f_{\mathrm{ext}}\\
\end{equation}
denotes the mechanical work performed per unit time and
\begin{equation}
\dot{Q} =-\frac{1}{TL}\int_0 ^L dx \int_0 ^T dt \,j \, \partial_x p\\
\end{equation}
denotes the interaction energy consumed per time.
The mean velocity and transport efficiency are shown in
Figs.~\ref{loadeff}\,(a) and (b).
\begin{figure} 
\includegraphics[width=0.8\hsize]{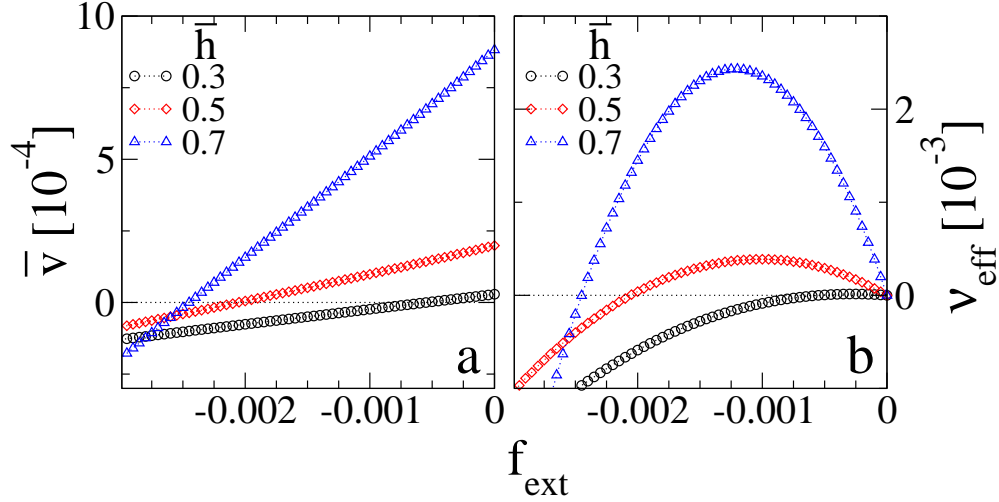}
\caption{The influence of an external load $f_{\mathrm{ext}}$. Shown are
the variations of (a) the mean velocity $\bar{v}$ 
and (b) the transport efficiency $\nu_{\mathrm{eff}}$
with $f_{\mathrm{ext}}$ for various mean film thicknesses $\bar{h}$ as indicated in the
legends.
The remaining parameters are as in Fig.~\ref{evol1} except for $T=500$.}
\mylab{loadeff} 
\end{figure}
The transport efficiency behaves non-monotonically. It increases for small loads,
reaches a maximum and decreases for higher loads. It becomes negative when the flow
reverses. Then the ratchet can no longer perform work against the external force.

Finally, we discuss the feasibility of a fluid ratchet employing 
spatially inhomogeneous electric fields by estimating the acting 
forces and relevant time and length scales for millimetric, microscopic and
nanoscopic fluidic systems. 
Typical material constants for thin films in capacitors are taken from 
\cite{EnSw00,Lin02,MPBT05} and references therein.
Applying a voltage of 100\,V over a gap of 2\,mm width, a 1\,mm film of 
silicon oil feels an electrostatic pressure of $p_{el}\approx0.1$N\,m$^{-2}$
corresponding approximately to the  curvature pressure in a drop of 1\,mm height with
$\theta_0=10\,^\circ$ equilibrium contact angle.
Thinner films can be influenced by a lower voltage \cite{Lin01}. 
For instance, for a 10\,$\mu$m oil film in a 
$20\,\mu$m gap for $U_0=10\,$V we find $p_{el}\approx10$\,N\,m$^{-2}$,
equivalent to the curvature pressure in a droplet of 10\,$\mu$m
height and $\theta_0=10\,^\circ$.
In principle, nanofluidic transport of polymer melts is also feasible.
For a 30\,nm film of liquid polystyrene in a 100\,nm gap which is
electrostatically patterned ($U_0=10$\,V) \cite{Lin02}, 
$p_{el}\approx10^4$N/m$^2$ equivalent to the curvature pressure for
a droplet of 30\,nm height and 300\,nm width.
Note, however, that for ultrathin films the disjoining pressure is of the  
same order of magnitude \cite{Isra92}.

Beside the length scales also the time scales are of uttermost importance.
The scale $\tau=3\gamma\eta/d\kappa^2_{el}$ reflects the relevant properties
responsible for the relaxation towards the flat film.  For a 5cS silicon oil
the estimate gives a timescale of 10 to $10^3$\,s, i.e.\ the viscous flow is
rather slow. Using, however, water films of thicknesses between 10\,$\mu$m
($U_0=10$V) and 1\,mm ($U_0=100$V) the time scale ranges from $10^{-5}$ to
$10^{-3}$\,s resulting in fast transport.

In conclusion, we have shown that a flashing ratchet produces a
macroscopic transport in a liquid film with a free surface on a feasible
time scale. There exist regimes of maximum transport selected by 
the spatial and temporal properties of the ratchet, which depend on the
characteristics of the thin film. For a spatial period of the ratchet
which is considerably larger than the length scale of the spinodal
instability of the flat film subjected to a homogeneous potential the
coarsening dynamics influences the transport in a non-trivial way.

\vspace{2ex}

This work was supported by the EU under grant MRTN-CT-2004-005728.
K.~J. was supported by grants from the CNES and the Humboldt-Foundation.


%
\end{document}